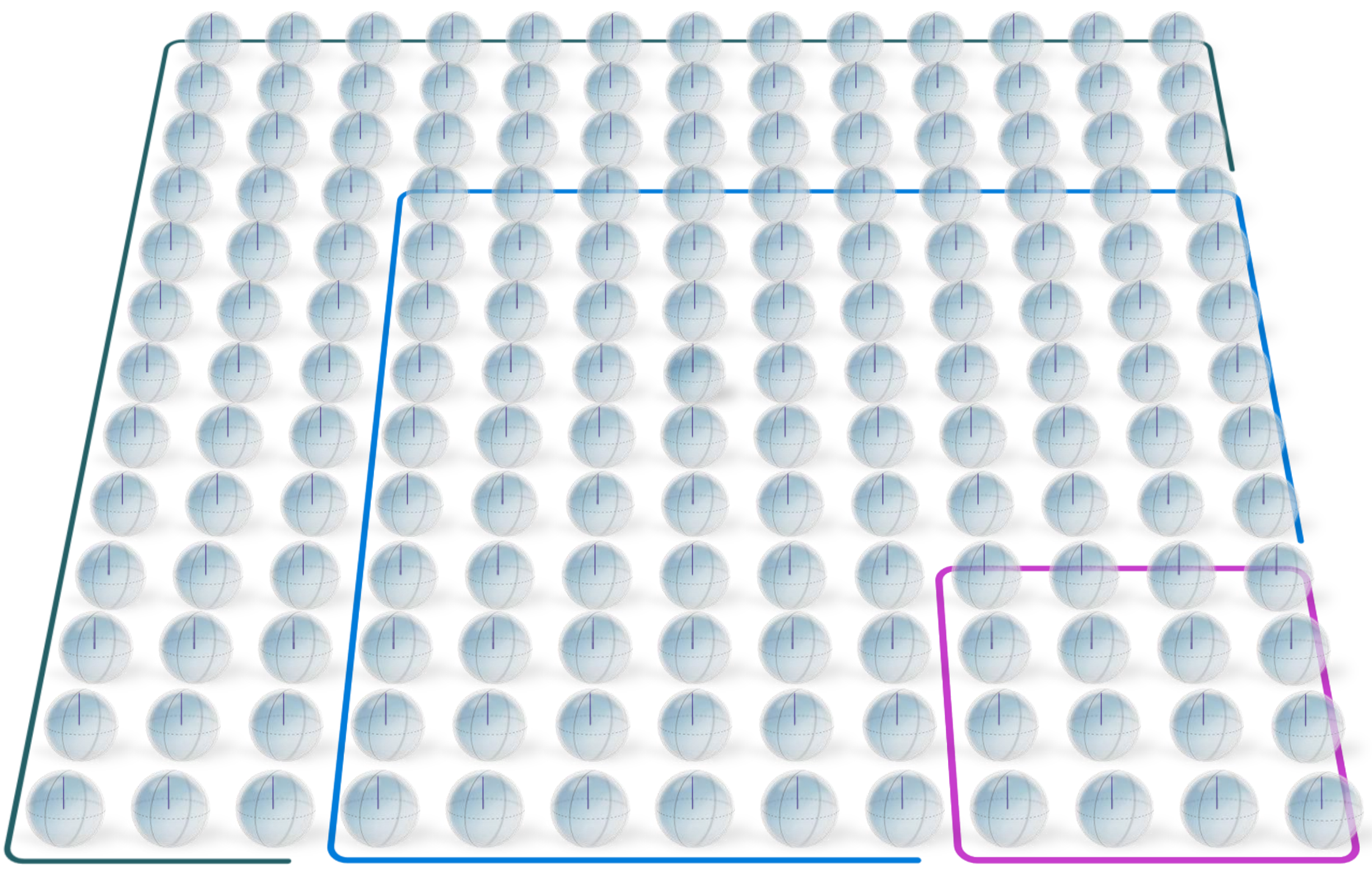

# Scalable logical qubits

Matthias Troyer, Microsoft Quantum
Chetan Nayak, Microsoft Quantum
John Martinis, QOLAB

## Abstract

Utility-scale quantum computing will require executing long, complex algorithms with end-to-end error rates far below what physical qubits can support directly. Error-corrected logical qubits are needed to achieve this goal. To make the progress on logical qubits measurable and comparable we introduce the definition of *scalable logical qubits*: logical qubits preserved for long computations by repeated quantum error correction, capable of fault-tolerant universal operations with low-latency real-time decoding and feedback, and replicable to the hundreds or thousands, as required by applications. We characterize these scalable logical qubits along four coupled dimensions: reliability, scale, capability, and performance, and discuss trade-offs among these dimensions.

# Introduction

The goal of quantum computing is to build machines that can routinely—rather than in one-off, hero demonstrations—solve valuable problems that are classically intractable. In this sense, the destination is utility-scale quantum computing, which involves computations that deliver clear practical value on real workloads. The most compelling targets are applications where quantum algorithms promise significant advantages in accuracy, speed, or achievable simulation scale [1]. Examples include simulating quantum systems, from simulating quantum physics models [2] to accurate simulations of molecules and materials [3], all the way to reaction pathways [4]; accelerating certain number-theoretic tasks and cryptanalysis [5]; and broader classes of problems that may become relevant as hardware and algorithms mature.

What these applications share is not only algorithmic promise, but demanding *resource requirements*: They require hundreds to thousands of high-quality qubits, sustained over circuits comprising billions of operations, with end-to-end failure probabilities small enough that the computation succeeds without prohibitive repetition. Concretely, we expect the “low end” of utility to begin around 100+ qubits with error rates on the order of $10^{-10}$, while high-end applications, such as large-scale chemistry and cryptanalysis, exceed 1000 qubits with error rates of $10^{-15}$ or better [6].

Physical qubits alone, even with improved physical reliability and combined with error mitigation, error detection, and post-selection, cannot support arbitrarily deep computations in a scalable way: As circuits grow, the probability of an error-free run decreases exponentially, and the number of required repetitions grows exponentially with computational complexity.

Reaching the low error rates that will enable utility scale demands more than better qubits, smart compilation, and post-processing; it demands *logical qubits* and *repeated quantum error correction* throughout program execution, so that errors are not merely detected but actively corrected, enabling computations of essentially unbounded duration [7].

# Logical qubits and quantum error correction

Early in the history of quantum computing it was unclear whether quantum information could be protected against noise at all. Peter Shor’s seminal work [8] showed that quantum error correction (QEC) is possible and the breakthrough was the proof of the threshold theorem stating that errors could be suppressed to arbitrarily low levels [9].

Similar to classical error correction, QEC protects information by encoding it redundantly across many physical qubits to form a single logical qubit. Because we cannot directly measure, compare, or copy an unknown quantum state, QEC instead extracts *error syndromes*: information about errors obtained by entangling the data qubits with auxiliary syndrome qubits and measuring the auxiliaries. In effect, syndrome measurements convert continuous physical noise into a stream of discrete errors that can be decoded into corrections—an error-correction mechanism that has no direct analogue in classical analog computers, where continuously valued state variables are inherently hard to stabilize against small perturbations.

Quantum error correction has come a long way since the initial code families such as CSS codes [10] and the surface code [11], and recently there has been a “Cambrian explosion” of new code developments especially around quantum low-density parity check (QLDPC) codes [12]. Combined with rapid developments in quantum hardware this puts utility scale quantum computing in sight for the next years.

## Scalable logical qubits for utility scale quantum computing

With the feasibility of quantum error correction, the foundation is laid for utility-scale quantum computing. The terminology we use to describe progress though has not kept pace with the requirements of utility-scale quantum computing. In particular, the term *logical qubit* is used today for a wide range of demonstrations that incorporate some elements of quantum error correction, and those definitions are often confusing or not clearly aligned with what is ultimately needed to run long, valuable algorithms. This makes it difficult to compare results across platforms and to assess where the growing number of logical qubit demonstrations [13] is on the roadmap from near-term milestones to the end goal. To address this, we define the term *scalable logical qubit* as the central concept of this paper.

**Definition:** A *scalable logical qubit* is a logical qubit with a path to meet the requirements for utility scale. It

(i) is sustained during computation by repeated quantum error correction,

(ii) supports a *fault-tolerant universal* set of logical operations, enabling measurement-conditioned control flow, through low-latency real-time error correction,

(iii) is an instance of a *code family* (and supporting quantum architecture) in which the logical error rate decreases predictably as more physical qubits are devoted to the logical qubit (e.g., by increasing code distance),

(iv) has a credible path to replication, i.e., to operating hundreds to thousands of such logical qubits while preserving logical performance. A scalable logical qubit should not require more than $O(N \log \epsilon^{-1})$ physical qubits to realize $N$ logical qubits with logical error rate $\epsilon$.

This goal-oriented definition of a scalable logical qubit provides a clear way to connect from today’s demonstration milestones to utility scale.

## Characterizing scalable logical qubits

We characterize a scalable logical qubit along four coupled dimensions—reliability, scale, capability, and performance—which together determine whether a platform can progress from small prototypes to systems that run demanding workloads.

- **Reliability:** logical error rates that initially are at least better than physical in an architecture and code family that has a path to logical error rates in the ~10-12–10-15 range, appropriate to the target application.
- **Scale:** a logical qubit architecture scaling hundreds to thousands of scalable logical qubits, with a system architecture that scales efficiently as qubit count increases — including modular/networked designs where applicable.
- **Capability:** universal, fault-tolerant operations with low-latency real-time decoding and feedback, enabling arbitrary programs and control flow.
- **Performance:** fast logical cycles and efficient gate synthesis so that wall-clock runtime and cost remain practical at scale.

There is no single path to these logical qubits; credible approaches must advance along all of these axes: reliability, scale, capability, and performance, and progress should be measured toward that goal and not along any individual metric.

**Capability** is the least obvious of the above, and we thus begin with a clear and consistent definition. A scalable logical qubit needs to ultimately provide capabilities spanning from state preparation and memory to universal computation with measurement-conditioned branching:

1. **State preparation and measurement:** prepare logical qubits in specified initial states and measure them with well-characterized error models.

2. **Memory:** store quantum information over extended durations, both in wall-clock time and relative to the native gate/measurement cycle time.

3. **Clifford operations:** implement a complete set of fault-tolerant Clifford gates (and corresponding measurements), synthesized from the platform's primitive operations.

4. **Universal (non-Clifford) operations:** extend beyond Clifford gates to a universal fault-tolerant gate set, typically via the preparation of magic-states [14].

5. **Measurement-conditioned control flow:** allow program execution to branch on measurement outcomes, requiring low-latency real-time decoding, feedforward, and (when needed) just-in-time gate synthesis.

These capabilities are familiar in spirit to DiVincenzo's criteria for physical-qubit quantum computing [15], but scalable logical qubits raise the bar: These primitives must work *under repeated error correction*, with fault-tolerant implementations, and with classical processing fast enough to stay in the feedback loop.

Framing capability explicitly helps to understand where a logical-qubit demonstration is on the path to an architecture that can support arbitrary algorithms as systems scale.

**Reliability:** A scalable logical qubit must achieve *logical error rates* that are better than the underlying physical error rates *under repeated error correction*. Equally important, it should come with a demonstrated *path to improvement*: the logical qubit is an instance of a code family (or architecture) in which increasing code distance (and corresponding resources) predictably reduces logical error rate, providing a credible route to the demanding $10^{-12}$–$10^{-15}$ regime required by utility-scale workloads.

**Scale:** The *number of logical qubits* that can be operated simultaneously, including scalable control and error correction. Importantly, one needs to consider how logical error rates and operation latencies change as the system scales: adding more logical qubits should not come at the expense of reduced functionality, degraded logical fidelity, or significantly increased correlated errors.

**Performance:** At utility-relevant problem sizes, additional performance metrics will become central and cost relevant, such as logical cycle times and gate latencies — not only the duration of one error-correction cycle, but the end-to-end latency to enact fault-tolerant logical operations. While initially speed may be the most relevant performance metric, ultimately it will be cost for reliable operations.

## Trade-offs

These metrics are tightly coupled: improving one often changes the achievable range of the others. Progress toward scalable logical qubits is thus shaped by trade-offs between fidelity, qubit count, cycle time, compilation overhead, and decoding latency. As a result, single headline numbers are rarely meaningful without context about the operating point and the constraints under which they were achieved. Such tradeoffs include

- **Qubit count versus fidelity:** With a fixed physical-qubit budget, one can allocate resources toward more logical qubits with modest improvements, or fewer logical qubits with substantially lower logical error rates. For scalability, logical-qubit count should be considered together with logical error rate and supported capability level.

- **Qubit count versus runtime:** Code choices that reduce the physical-to-logical qubit ratio (space) may increase the number of logical cycles needed for fault-tolerant gates (time), and vice versa. A favorable operating point depends on the target workload and which resource—hardware qubits or runtime—is more constrained.

- **Decoder accuracy versus latency:** Decoding is the classical inference step that turns streaming syndrome data into corrections and feed-forward decisions. Decoders vary in both logical performance and runtime: some achieve higher correction accuracy at higher computational cost; others trade accuracy for speed. The goal is an operating point that enables reliable *real-time* decoding and feedback without materially degrading logical error rates.

- **Code optimization versus compile time and runtime overhead:** More aggressive compilation and gate synthesis can reduce runtime or qubit count required for an application. This includes the number of magic states, routing overhead, and peak ancilla usage. The cost is increased classical work: either longer offline compile times, or (if performed at runtime) tighter real-time latency and integration constraints in the decode/feedforward loop.

## Additional performance levers beyond error correction

Repeated quantum error correction is the enabling ingredient for scalable logical qubits, but it is not the only lever for improving end-to-end performance. Higher physical-qubit fidelities, selective error detection with post-selection, and error-mitigation techniques applied at the logical level can reduce overheads alongside QEC.

**Improving physical-qubit fidelity** remains foundational. Lower physical error rates increase the margin to threshold, reduce the code distance required to reach a given logical error rate, and can decrease both space overhead (fewer physical qubits per logical qubit) and time overhead (shorter or fewer correction cycles for a target reliability).

**Error detection and post-selection** can also be powerful. By post-selecting on "no-detected-error" events, experiments can demonstrate improved overall effective fidelities [16] of an application or used as a subroutine produce higher quality resource states [17]. However, post-selection at the application level by itself does not scale to long computations: As circuit depth grows, the probability of an error-free run typically falls rapidly, leading to a prohibitive growth in the number of required attempts.

**Error mitigation**, including techniques such as extrapolation, symmetry verification, and probabilistic error cancellation, can further improve accuracy when applied on top of error-corrected logical operations [18]. Used judiciously, mitigation can reduce the effective error seen by an application and relax the burden on the outermost logical error targets. Nevertheless, mitigation methods typically incur additional sampling overhead and rely on assumptions that become harder to satisfy at very large scales. As a result, they are again used to complement rather than replace repeated error correction and scalable logical qubits.

## Performance and Cost

While less relevant for initially reaching application scale, cost ultimately determines whether a quantum computation is useful in practice. The Defense Advanced Research Projects Agency (DARPA) defines utility-scale quantum computing as the point where the value of a computation exceeds the cost to perform it and adopts this as an operational lens for evaluating architectures and roadmaps [19].

For fault-tolerant systems, cost is driven by a combination of:

- Wall-clock runtime determined logical cycle time and the number of cycles required by the algorithm and gate synthesis.

- Error-correction overhead (physical-to-logical qubit ratio and additional time/space needed to reach a target logical error rate),

- Low-latency real-time decoding and feedforward, which consumes classical compute and constrains system design, and

- The number of repetitions of the quantum computation required to achieve a desired end-to-end success probability.

The ultimate relevant question is not just “how many qubits,” but how much reliable computation per unit time and hardware can the system deliver and at what cost.

# Conclusion

Utility-scale quantum computing requires running long, complex programs with extremely low end-to-end failure probability. This shifts the central unit of progress from physical qubits to logical qubits. Because “logical qubit” is used broadly across today’s prototypes, we have proposed a tighter, goal-oriented definition *of scalable logical qubits* and a practical way to track progress.

The four dimensions reliability, scale, capability, and performance provide a common language for assessing how far along a given approach is to enabling real applications, and what must improve next. Ultimately, framed through cost, the goal is clear: deliver more *reliable computation* per unit time and hardware.

Questions to be asked of early demonstrations on this path include:

- How many cycles of repeated error correction have been demonstrated and what is required to do more?
- What capabilities have been demonstrated: state preparation, memory, Clifford gates, universal gates and real-time feedback? What is still required and how can it be achieved?
- What are the logical error rates? Are they better than physical, and how can they be further decreased by increasing the number of physical qubits used for the logical qubit?
- How many logical qubits have been realized? How can their number be increased further while keeping the error rates as they are?
- What is the control architecture to support the number and quality of qubits required for utility-scale applications?

Assessing progress along the envelope of the four dimensions of reliability, scale, capability, and performance, rather than quoting individual numbers, will help track progress and clearly identify advances made along one or multiple dimensions. With shared definitions and metrics, the community can compare approaches meaningfully and align near-term milestones [13] with the long-term destination.

---

1 T. Häner, D. S. Steiger, T. Hoefler, and M. Troyer. 2021. Distributed Quantum Computing with QMPI. In Proceedings of the International Conference for High Performance Computing, Networking, Storage and Analysis (SC '21). Association for Computing Machinery, New York, NY, USA, Article 16, 1–13.

2 A. J. Daley, I. Bloch, C. Kokail, S. Flannigan, N. Pearson, M. Troyer, and P. Zoller, "Practical quantum advantage in quantum simulation," *Nature* 607, 667–676 (2022). https://doi.org/10.1038/s41586-022-04940-6

3 R. P. Feynman, Simulating physics with computers, *International Journal of Theoretical Physics*, 21 (1982), 467–488, 1982, https://doi.org/10.1007/BF02650179; Lloyd, S. (1996). Universal quantum simulators. *Science*, 273(5278), 1073–1078. https://doi.org/10.1126/science.273.5278.1073; Bela Bauer, Sergey Bravyi, Mario Motta, and Garnet Kin-Lic Chan. 2020. Quantum Algorithms for Quantum Chemistry and Quantum Materials Science. *Chemical Reviews* 120, 22 (2020), 12685–12717. https://doi.org/10.1021/acs.chemrev.9b00829; Bela Bauer, Sergey Bravyi, Mario Motta, and Garnet Kin-Lic Chan. 2020. Quantum Algorithms for Quantum Chemistry and Quantum Materials Science. Chemical Reviews 120, 22 (2020), 12685–12717. https://doi.org/10.1021/acs.chemrev.9b00829

4 M. Reiher, N. Wiebe, K. M. Svore, D. Wecker, and M. Troyer, "Elucidating reaction mechanisms on quantum computers," *Proceedings of the National Academy of Sciences*. 114 (2017) 7555–7560.

5 Peter W. Shor. 1997. Polynomial-Time Algorithms for Prime Factorization and Discrete Logarithms on a Quantum Computer. *SIAM Journal on Computing* 26, 5 (October 1997), 1484–1509. https://doi.org/10.1137/S0097539795293172

6 M. E. Beverland, P. Murali, M. Troyer, K.M. Svore, T. Hoefler, V. Kliuchnikov, G.H. Low, M. Soeken, A, Sundaram, and A, Vaschillo, "Assessing requirements to scale to practical quantum advantage," *arXiv preprint* arXiv:2211.07629, 2022. https://arxiv.org/abs/2211.07629; Craig Gidney, "How to factor 2048 bit RSA integers with less than a million noisy qubits," *arXiv preprint* arXiv:2505.15917, 2025. https://arxiv.org/abs/2505.15917

7 Nielsen & Chuang, Quantum Computation and Quantum Information (2000).

8 P.W. Shor, Scheme for reducing decoherence in quantum computer memory, Physical Review A 52, R2493-R2496 (1995).

9 P. W. Shor, "Fault-Tolerant Quantum Computation," in *Proceedings of the 37th Annual Symposium on Foundations of Computer Science*, 56–65 (1996), arXiv:quant-ph/9605011; D. Aharonov and M. Ben-Or, "Fault-Tolerant Quantum Computation with Constant Error Rate," *SIAM Journal on Computing* 38, 1207–1282 (2008), arXiv:quant-ph/9906129; E. Knill, R. Laflamme, and W. H. Zurek, "Threshold Accuracy for Quantum Computation," arXiv:quant-ph/9610011 (1996); A. Yu. Kitaev, "Quantum Computations: Algorithms and Error Correction," *Russian Mathematical Surveys* 52, 1191–1249 (1997).

10 A. R. Calderbank and P. W. Shor, "Good Quantum Error-Correcting Codes Exist," *Physical Review A* 54, 1098–1105 (1996), arXiv:quant-ph/9512032, doi:10.1103/PhysRevA.54.1098; A. M. Steane, "Multiple-

---

19 https://www.darpa.mil/research/programs/underexplored-systems-for-utility-scale-quantum-computing